\documentclass[prl,aps,twocolumn]{revtex4}
\usepackage{epsfig}
\usepackage{bm}
\newcommand{\B}[1]{{\bm{#1}}}

\usepackage[latin1]{inputenc}
\newcommand{\beq}{\begin{equation}}
\newcommand{\eeq}{\end{equation}}
\newcommand{\bea}{\begin{eqnarray}}
\newcommand{\eea}{\end{eqnarray}}
\begin{document}
\title{Disentangling Scaling Properties in Anisotropic Fracture}
\author{Eran Bouchbinder, Itamar Procaccia and Shani Sela}

\affiliation{Dept. of Chemical Physics, The Weizmann Institute of
Science, Rehovot 76100, Israel}

\begin{abstract}
Structure functions of rough fracture surfaces in isotropic
materials exhibit complicated scaling properties due to the broken
isotropy in the fracture plane generated by a preferred propagation direction.
Decomposing the structure
functions into the even order irreducible representations of the SO(2)
symmetry group (indexed by $m=0,2,4\dots$) results in a lucid and
quickly convergent description. The scaling exponent of the
isotropic sector ($m=0$) dominates at small length scales. One can
reconstruct the anisotropic structure functions using only the
isotropic and the first non vanishing anisotropic sector ($m=2$) (or
at most the next one ($m=4$)). The scaling exponent of the isotropic
sector should be observed in a proposed, yet unperformed,
experiment.
\end{abstract}

\maketitle

\begin{figure}
\begin{center}
\epsfysize=4.0 truecm \epsfbox{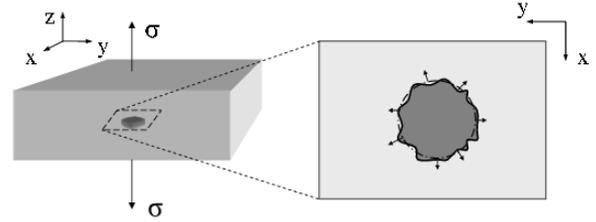} \caption{Sketch of a
hypothetical fracture experiments arranged to allow a crack to
develop in an isotropic fashion, i.e. with all angles $\theta$ being
statistically equivalent.On the left, the full three dimensional
experiment is shown, where the tensile axis is along $z$ and a
circular cavity is in the $xy$ plane. On the right, a magnified
version of the circular cavity in the $xy$ plane is shown.}
\label{experiment}
\end{center}
\end{figure}

{\bf Introduction}: Imagine a fracture experiment in which an
initial circular cavity is made to propagate by a tensile load such
that the crack edge remains circular on the average, without any
preferred propagation direction in the plane normal to the load, see
Fig. \ref{experiment}. From the point of view of the scaling
properties of the rough fracture surface that is left behind the
advancing crack, such an experiment is the analogue of homogenous
and isotropic turbulence in nonlinear fluid mechanics. Normal
experiments in both fracture and turbulence involve symmetry
breaking; the boundary conditions introduce anisotropy, making the
discussion of scaling properties non-trivial. In turbulence it was
shown how to disentangle the anisotropic contributions from the
isotropic one by projecting the measured correlation and structure
functions on the irreducible representations of the SO(3) symmetry
group \cite{05BP}. The scaling phenomena seen in the isotropic
sector of anisotropic experiments are identical to those expected in
the hypothetical experiment of homogenous and isotropic turbulence.
The aim of this Letter is to introduce a similar concept to the
field of fracture: we will show that decomposing the height-height
structure functions of fracture surfaces into the irreducible
representations of the SO(2) symmetry group results in a
simplification  and rationalization of the scaling properties that
is not totally dissimilar to the one obtained in turbulence. The
scaling properties of the isotropic sector should be observable in
principle in an experiment like the one shown in Fig.
\ref{experiment}, which contrary to turbulence may be performed in
reality.

The pioneering experimental work described in Ref. \cite{84MPP} drew
attention to the fact that fracture surfaces are  graphs in 2+1
dimensions when the broken sample is three dimensional. This initial
insight was followed up by a considerable number of works
\cite{97Bou} that focused on the scaling properties of such graphs
under affine transformations. In 2+1 dimensions one denotes the
graph as $h(\B r)$ and considers the structure function $S_2(\B
\ell)$,
\begin{equation}
 S_2(\B \ell)\equiv \langle (h(\B r+\B \ell)-h(\B r))^2\rangle \ , \label{S2}
 \end{equation}
 where angular brackets denote an average over all $\B r$. Initially no attention was
 paid to the fact that the isotropy in the fracture plane is broken due to initial conditions that lead to a
preferred propagation direction,
 and the statement of \cite{84MPP} was that the structure function
 is a homogeneous function of its arguments,
$S_2(\lambda \B \ell)\sim \lambda^{\zeta^{(2)}} S_2(\B \ell) \ .$
 In fact such a statement is tenable only if the fracture process {\em and} the material are  isotropic. Usually the crack
 propagates predominantly in one direction (say  $\B {\hat x}$) and the vector $\B \ell$ defines an
 angle $\theta$ with respect to  $\B {\hat x}$, $\theta =\cos^{-1} (\B {\hat x}\cdot \B {\hat \ell})$. There is no reason
 why the scaling exponent $\zeta^{(2)}$, if it exists at all, should not depend on the angle $\theta$.
 Indeed, in the later work that followed \cite{84MPP} this problem was recognized and scaling
 exponents were sought for one dimensional cuts through $ S_2( \B \ell)$, typically parallel
 and orthogonal to the direction of the crack propagation. Besides the obvious meaning of
 `parallel' and `orthogonal' to  $\B {\hat x}$, no reason was ever given why these particular directions
 are expected to provide clean scaling properties. We argue below that
 in general such one dimensional cuts exhibit a mixture of scaling exponents with amplitudes
 that depend on the angle $\theta$, where $\theta=0$ and $\theta=\pi/2$ are not
 special.\\
 {\bf Approach}: Given an experimental surface $h(\B r)$ we first compute the second order
structure function defined in Eq. (\ref{S2}). The vector $\B \ell$
is associated with a norm $\ell$ and an angle $\theta$.
 By construction, the second order structure function is symmetric under $\theta \to \theta+\pi$.
 Accordingly, decomposing the structure functions into the irreducible representations of the SO(2) symmetry group results in summations over even indices only:
 \begin{equation}
 S_2(\ell , \theta) = \sum_{m =\; -\infty}^{\infty} a_{2m}(\ell)
e^{i 2 m \theta} \ . \label{decompose}
\end{equation}
Such a decomposition is deemed useful when each of the scalar functions $a_{2m}(\ell)$ is
itself a homogeneous function of its argument, characterized by an $m$ dependent exponent:
    \begin{equation}
        |a_{2m} (\lambda \ell )| \sim \lambda^{\zeta^{(2)}_{2m} }|a_{2m}(\ell)| \ ,
    \end{equation}
where $|\cdot |$ stands for the norm of a complex number. For an
isotropic fracture in an isotropic medium we expect $a_{2m}(\ell)=0$
for all $m\ne 0$. In usual mode I experiments in which the crack
propagates along the $\B {\hat x}$ direction and the tensile load is
in the normal direction, there should be the same physics along
lines with angles $\theta$ and $-\theta$. This invariance under
$\theta\to -\theta$ implies that the arguments of all
$a_{2m}(\ell)\ne 0$ should be 0 or $\pi$. In reality this invariance
might not hold when the fracture process breaks the symmetry
dynamically; see bellow for an example.

\begin{figure}
\begin{center}
\epsfysize=5. truecm \epsfbox{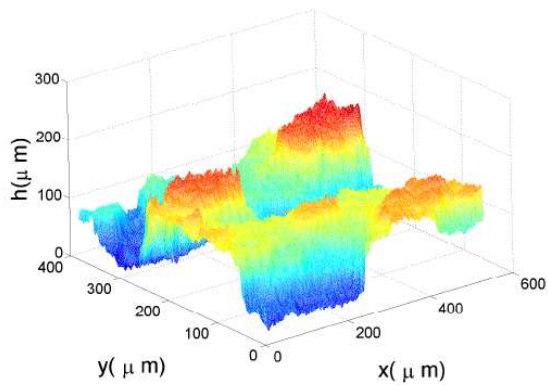} \caption{The raw fracture
surface of the aluminum alloy obtained in Ref. \cite{98AB}.}
 \label{raw}
\end{center}
\end{figure}

\begin{figure}
\begin{center}
\epsfysize=5 truecm \epsfbox{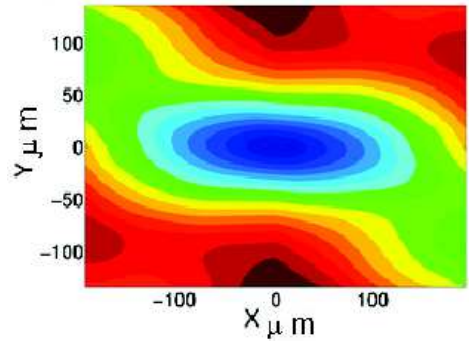}
\caption{Contour plot of the second order structure
function of the surface shown in Fig. \ref{raw}.}
 \label{FigS2}
\end{center}
\end{figure}

\begin{figure}
\begin{center}
 \epsfxsize=6.5 truecm \epsfbox{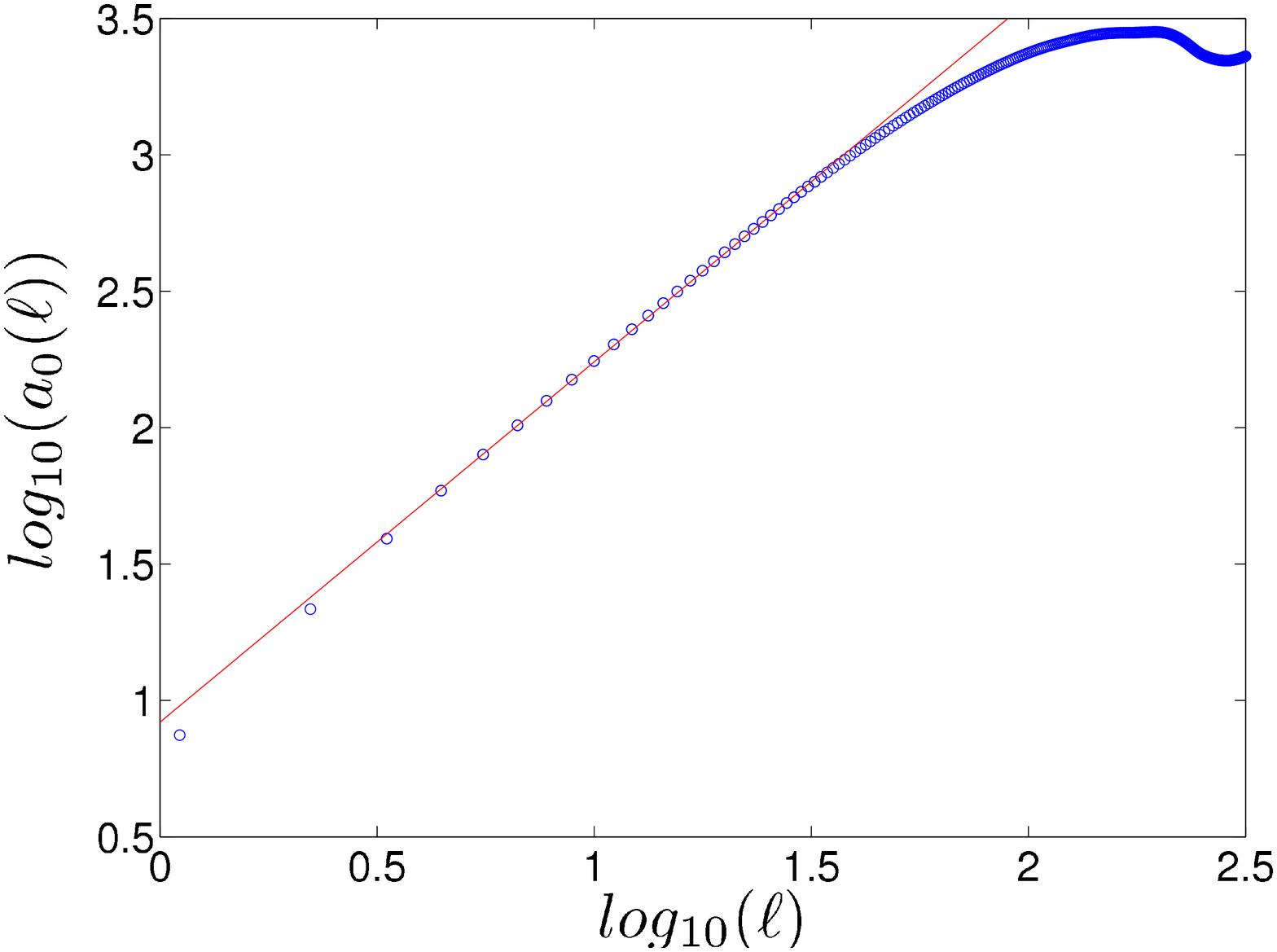}
 \epsfxsize=6.5 truecm \epsfbox{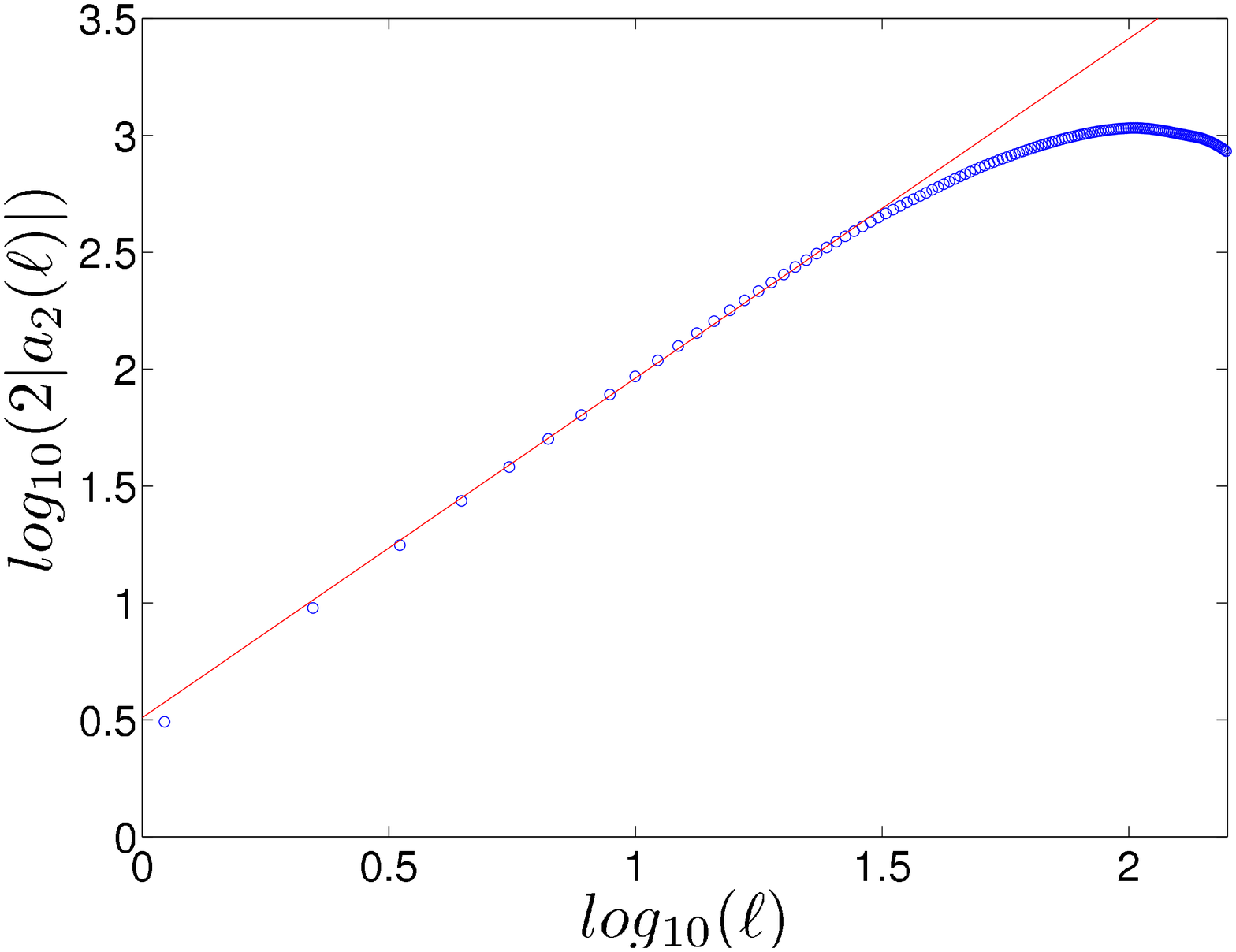}
 \epsfxsize=6.5 truecm \epsfbox{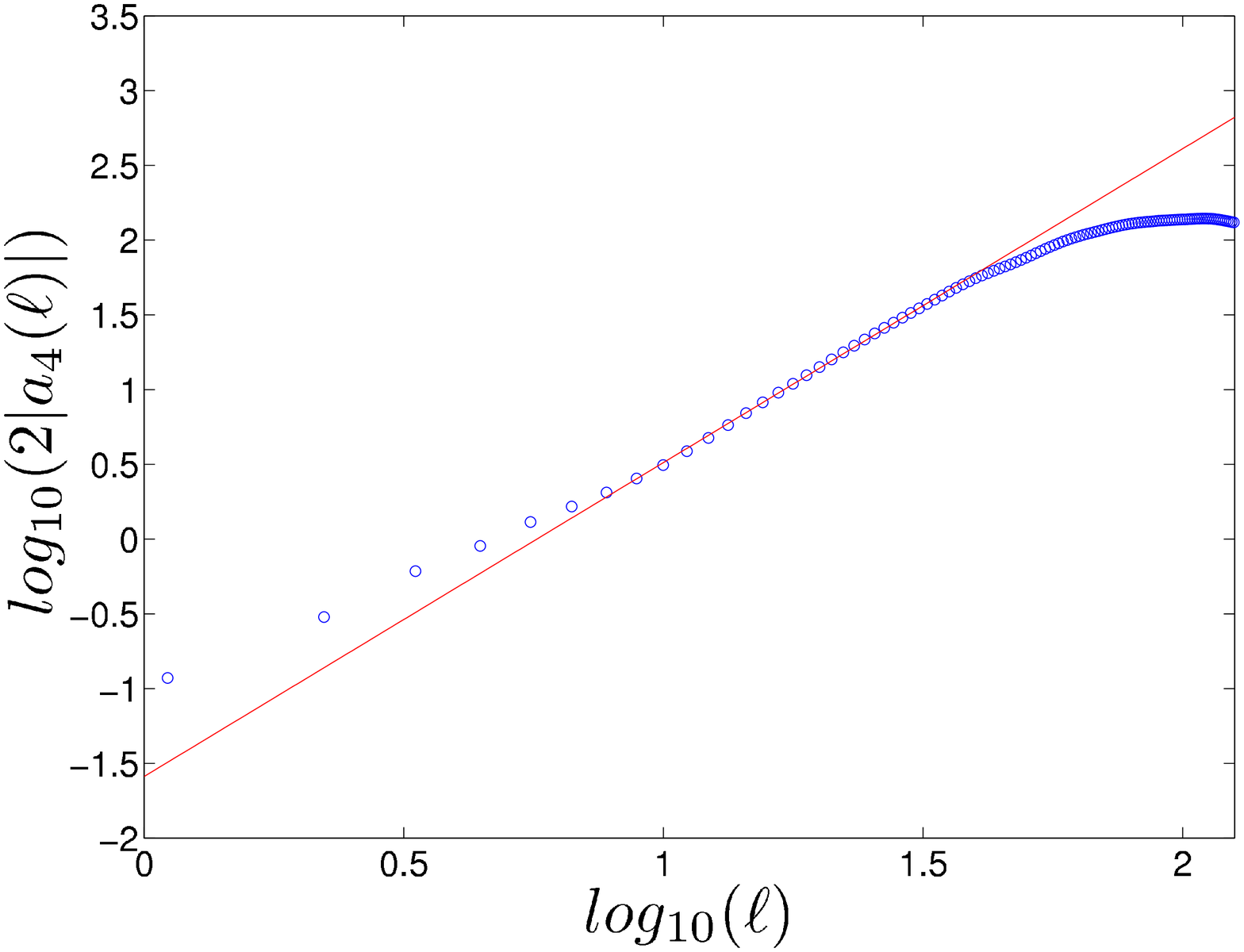}
\caption{Log-log plots of the amplitudes $a_0(\ell)$, $2|a_2(\ell)|$
and $2|a_4(\ell)|$ vs. $\ell$ for the aluminum alloy.}
 \label{alua4}
\end{center}
\end{figure}

{\bf Example: aluminum alloy -}  Our first example was obtained
\cite{98AB} from a compact tension specimen made of 7475
aluminum alloy first precracked in fatigue and then broken under
tension in mode I. The raw fracture surface and the second order
structure function computed from it are shown in Figs. \ref{raw} and
\ref{FigS2} respectively.
One sees the anisotropy of $S_2(\ell)$ with the naked eye. To
quantitatively characterize this anisotropy, the structure function
was decomposed as in Eq. (\ref{decompose}). The log-log plots of
$a_0(\ell)$, $2|a_2(\ell)|$ and $2|a_4(\ell)|$ are exhibited in Fig.
\ref{alua4}.
By performing linear fit of the relevant range in the log-log plots we find the following exponents
\begin{equation}
\zeta^{(2)}_0 \!\!= 1.32 \pm 0.07 \  ,  \zeta^{(2)}_2\!\! = 1.45 \pm 0.08 \ , \zeta^{(2)}_4 \!\!= 2.1\pm 0.1\ . \label{metzeta}
\end{equation}
The implication is that at smaller length-scales the smaller exponent $\zeta^{(2)}_0$
should be dominant and vice versa. Indeed, examining again the contour plot in Fig. \ref{FigS2} one
observes that at small scales the contours tend to ellipses of smaller eccentricity, whereas at larger scale the contours are ellipses with increasing eccentricity.

\begin{figure}
\epsfxsize=6 truecm \epsfbox{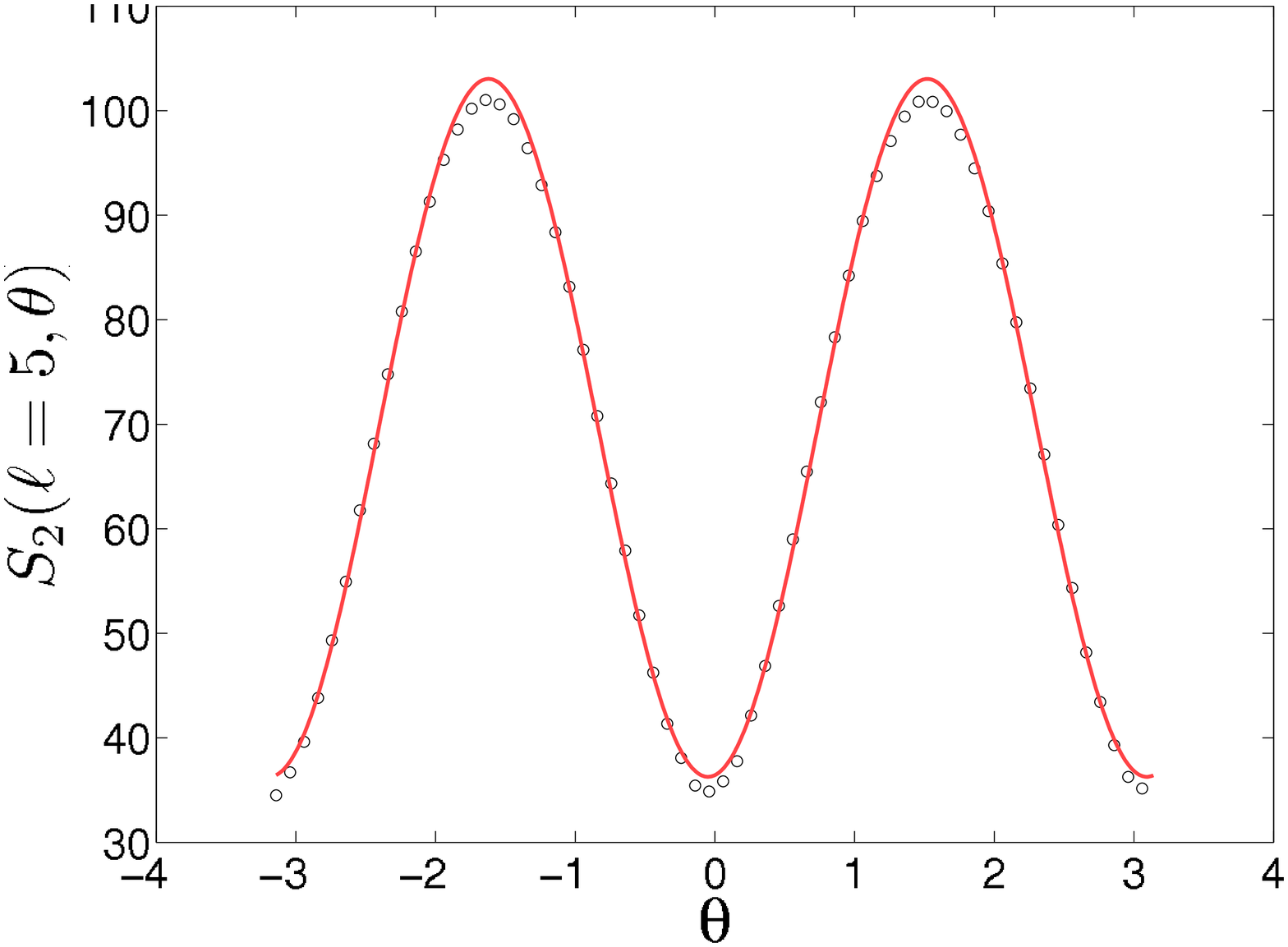} \epsfxsize=6
truecm \epsfbox{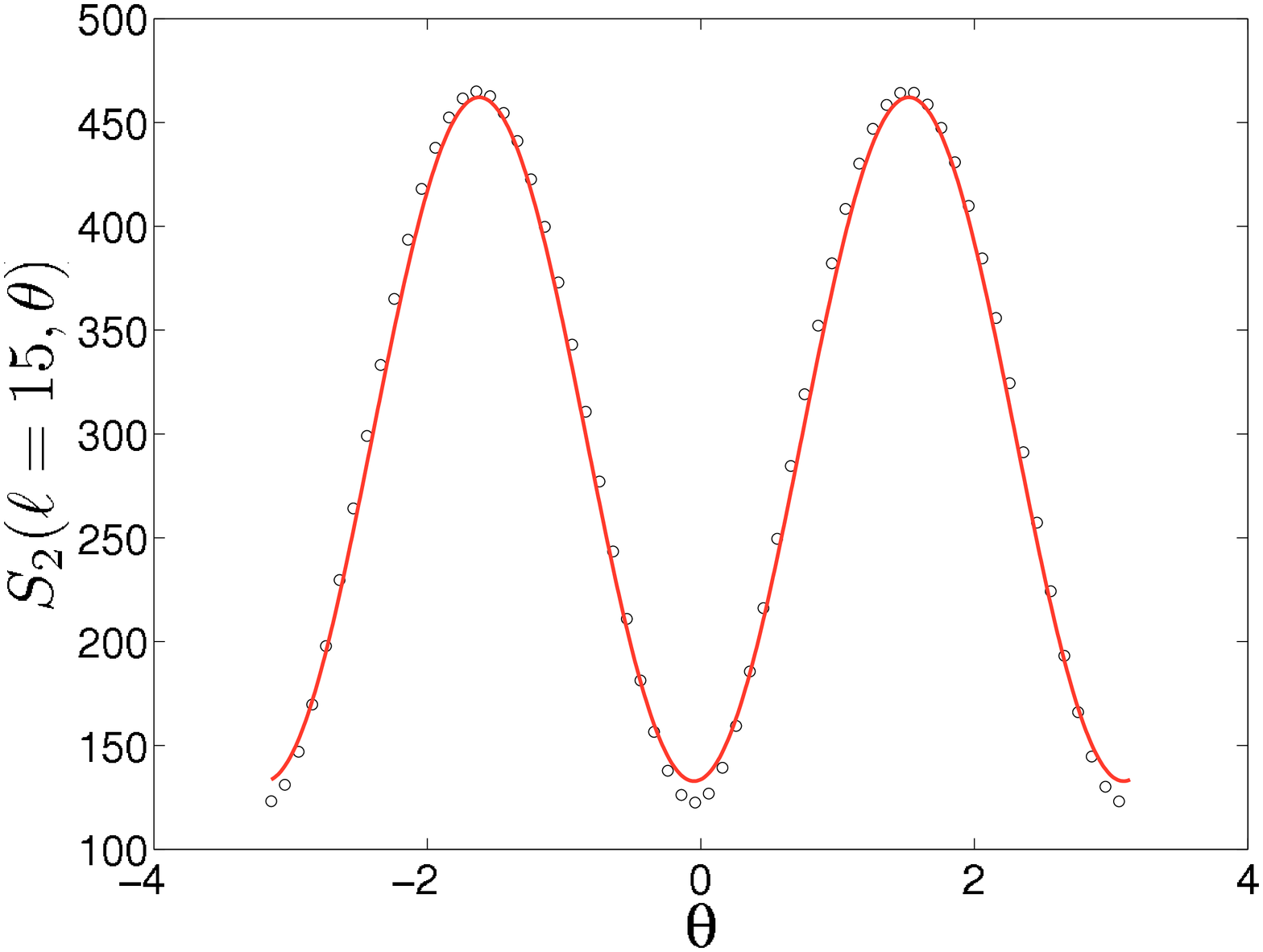} \epsfxsize=6 truecm \epsfbox{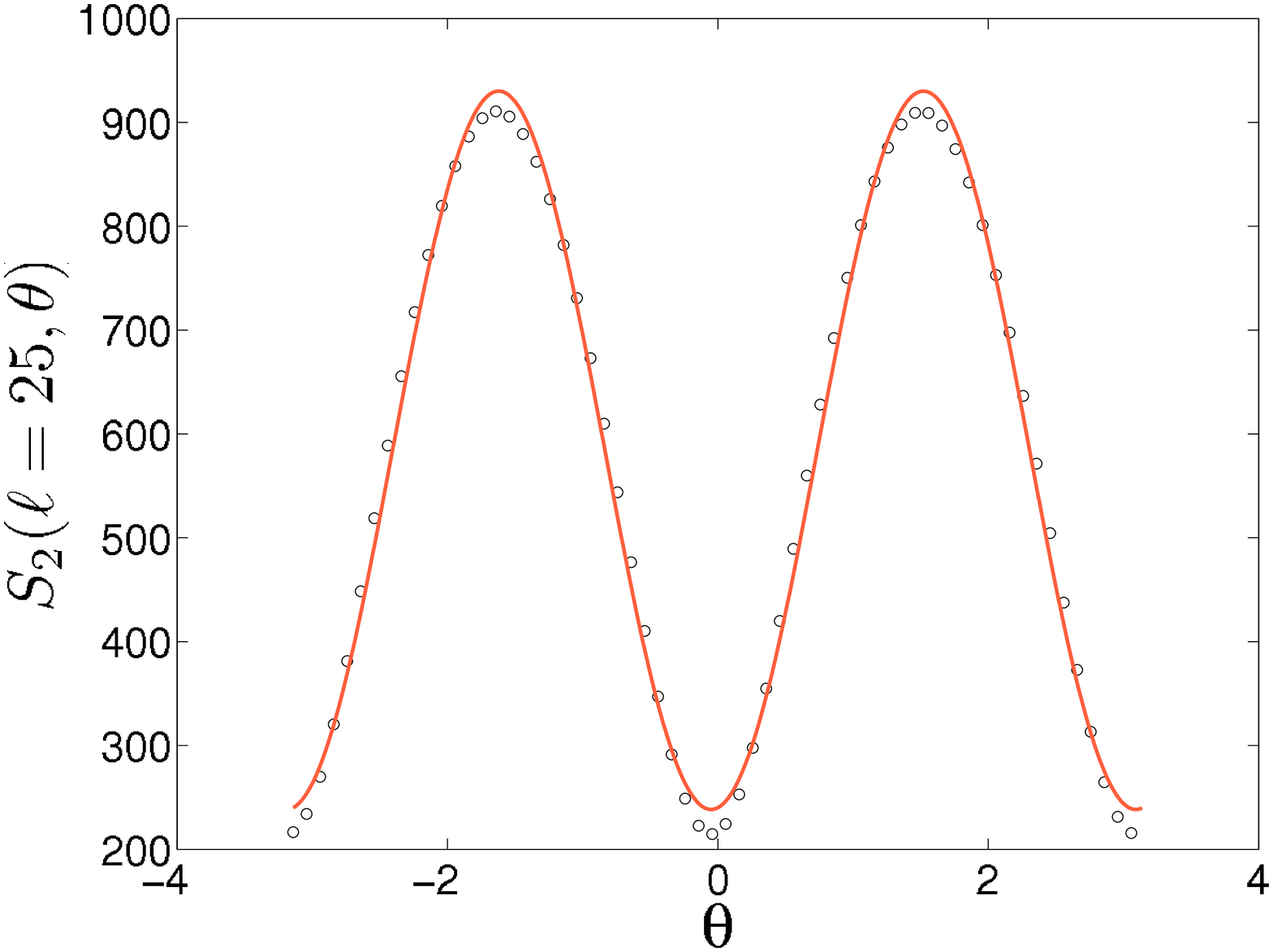}
\caption{The experimental $S_2(\ell, \theta)$ for the aluminum alloy
(circles) and the representation Eq. (\ref{represent}) (line), for
$\ell=5$, 15 and 25$\mu m$.}\label{compare}
\end{figure}

\begin{figure}
\epsfxsize=6 truecm \epsfbox{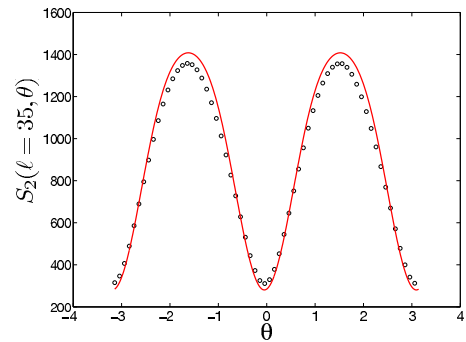} \caption{The experimental
$S_2(\ell, \theta)$ for the aluminum alloy (circles) and the SO(2)
expansion up to the third even order irreducible representation
(line), for $\ell=35\mu m$.} \label{35}
\end{figure}

The crucial test of this approach is whether one can reconstruct the structure function $S_2(\ell,\theta)$
using the functional form of the irreducible representation and a minimal number of parameters. Indeed,
at smaller values of $\ell$ the first two irreducible representations suffice. Writing
\begin{equation}
S_2( \ell, \theta) \approx 8.30~ \ell^{1.32}  +3.22 ~ \ell^{1.45} cos(2\theta + \pi) \ , \label{represent}
\end{equation}
we compare in Fig.\ref{compare} the experimental data to Eq.
(\ref{represent}) for $\ell=5$, 15 and 25$\mu m$. The excellent fit
is obvious. In fact, with four parameters (two amplitudes and two
exponents) we can represent the structure function to within 1\%  in
$L^2$ norm as long as $\ell\le 30\mu m$. For larger values of $\ell$
the agreement decreases, and we need to employ the next irreducible
representation. Adding $0.026~\ell^{2.1}\cos(4\theta+\pi)$, we find
the fit shown in Fig. \ref{35} for $\ell=35\mu m$. Beyond these
values the power-laws fits lose their credibility for this
experimental data set.

\begin{figure}
\epsfxsize=6 truecm \epsfbox{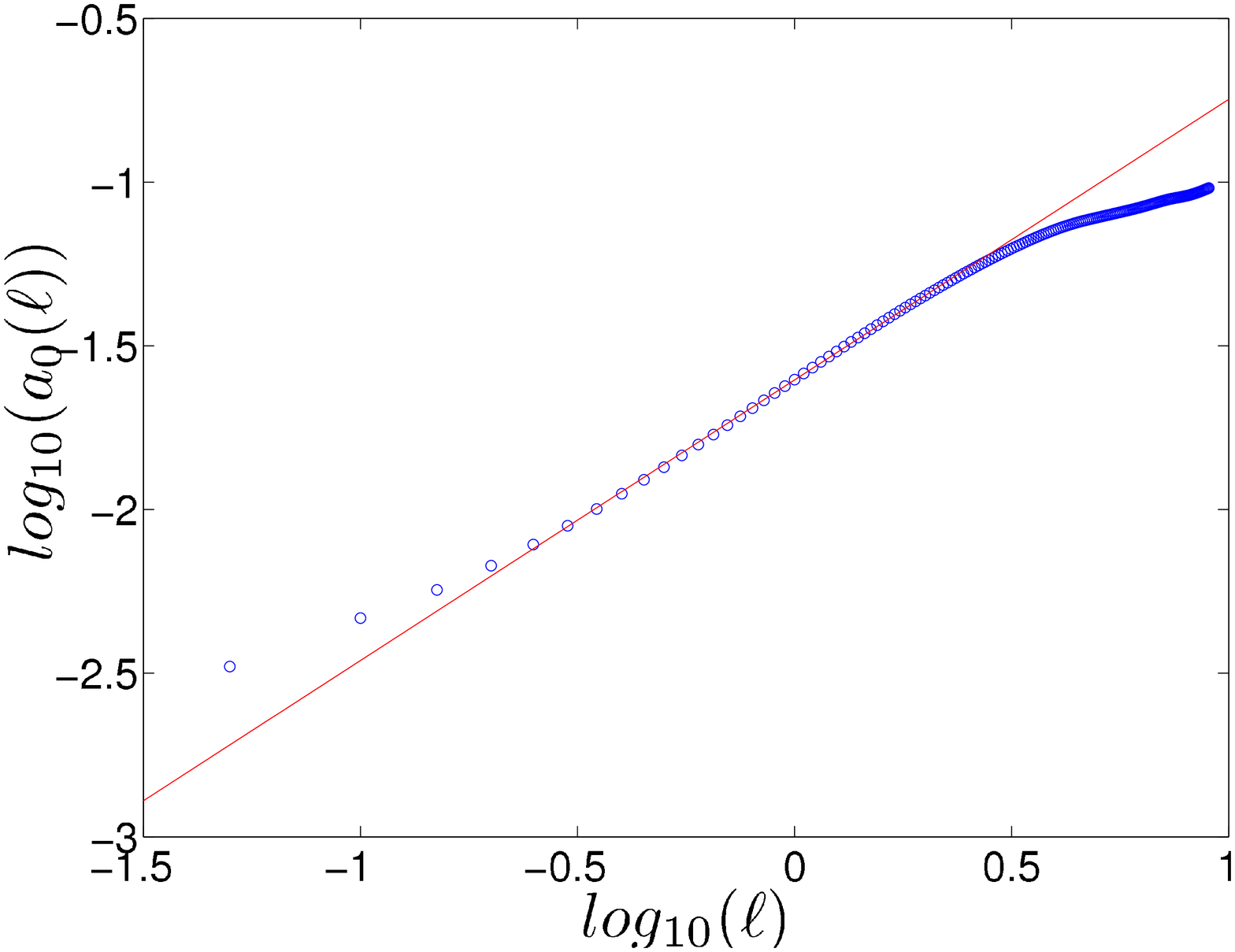}
\epsfxsize=6 truecm \epsfbox{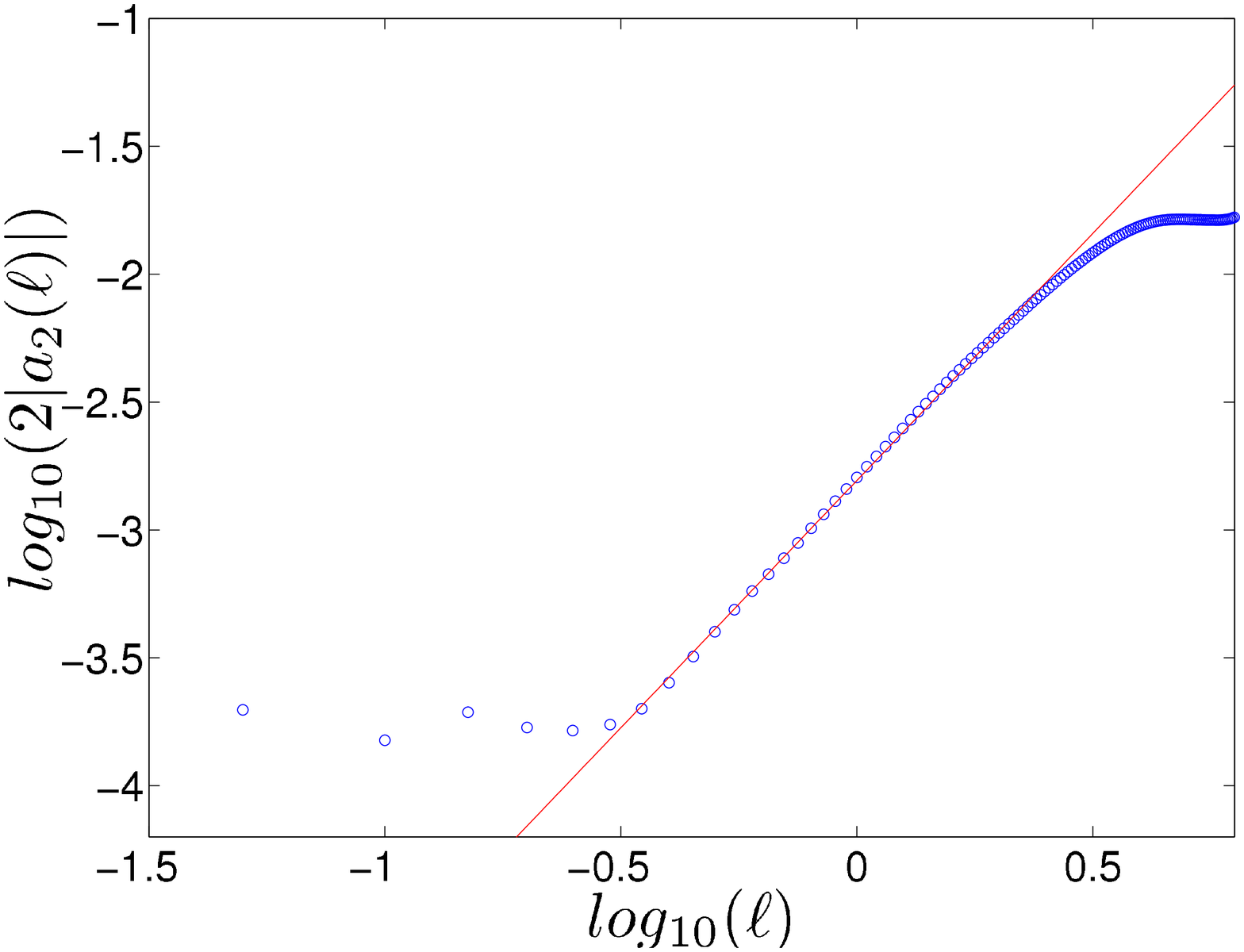}
\caption{Log-log plots of the amplitudes $a_0(\ell)$ and
$2|a_2(\ell)|$ for the artificial rock.} \label{a0a2rock}
\end{figure}

\begin{figure}
\epsfxsize=6 truecm \epsfbox{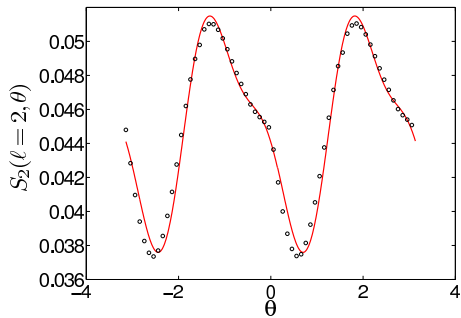}
\caption{Comparison of the experimental $S_2(\ell, \theta)$ (circles) and the SO(2) expansion up to the third even irreducible representation (line) for the artificial rock with $\ell=2 mm$.}
\label{r2}
\end{figure}

{\bf Example: artificial rock-}  The second example was obtained
from the dynamic fracture of artificial rocks produced from carbonatic
aggregates cemented by epoxy \cite{05Sag}. The samples are plates of
size $400 \times 400\times 9$ mm, and the fracture surface was
measured using a scanning laser profilometer. The analysis of the
experimental data follows verbatim the first example. The plots of
$a_0(\ell)$ and $2|a_2(\ell)|$ are shown in Fig. \ref{a0a2rock}.
Fitting the plots we find
\begin{equation}
\zeta^{(2)}_0 \!\!= 0.86 \pm 0.05\ , \zeta^{(2)}_2\!\! = 1.93 \pm
0.05\ , \zeta^{(2)}_4 \!\!= 1.93\pm 0.1\ . \label{rockzeta}
\end{equation}
Two comments are in order. First, one should notice the non-universality of the scaling exponents
as compared with the previous example (\ref{metzeta}). This is to be expected when comparing
a quasi-static and a rapid fracture experiments. Second, the present surface does {\em not}
satisfy a $\theta\to -\theta$ symmetry. This results from the dynamic instability that leads to
side branches that leave directed traces on the fracture surface, destroying the $\theta\to -\theta$
symmetry  \cite{05Sag}.
Due to the lack of symmetry the amplitudes of the coefficient $a_m$ can take any phase, not
constrained to 0 or $\pi$ as required by the $\theta\to -\theta$ symmetry. The lack of symmetry
is clearly obvious in the reconstruction of the structure function from the irreducible representations.
We point out the advantage of the present approach in clearly revealing, through
the lack of symmetry,  some aspects of the
underlying physical process generating the fracture surface.
In Fig. \ref{r2} we compare the experimental values of $S_2(\ell,\theta)$, for $\ell=2~ mm$, to the expansion
\begin{eqnarray}
S_2( \ell, \theta)& \approx &0.025~ \ell^{0.86}  +0.0016 ~ \ell^{1.93} \cos(2\theta + 2.09) \nonumber\\
&+& 5.4\times 10^{-4} ~\ell ^{1.93} \cos(4\theta - 0.17) \ , \label{represent2}
\end{eqnarray}
The fit is satisfactory and the asymmetry in $\theta$ is obvious.

Taking the present two examples as representative, it appears that the SO(2) decomposition
extracts pure scaling behavior in each sector, but that the scaling exponents are not universal,
at least in the two examples discussed here. Considering cuts in $S_2(\ell,\theta)$ along the
$\theta=0$ and $\theta=\pi$ directions, the present approach predicts a mixture of
scaling exponents rather than pure power laws. We verified that plotting the expansion (\ref{represent})
for $\theta=0$ and $\theta=\pi/2$ in log-log plots we can reproduce the {\em apparent} scaling
laws obtained by longitudinal and transverse cuts.
Finally, one should point out that the SO(2) decomposition is not expected to yield
satisfactory results when the material itself is strongly anisotropic. As an example we considered fracture
surfaces in wood. This is clearly an anisotropic medium due to the fiber structure, and indeed
we found that  along and across the fiber structure the scaling behavior appears credible,
whereas the SO(2) decomposition failed altogether to reveal clean scaling properties
in any sector.

{\bf Summary}: we propose that materials which can be fractured in
an isotropic fashion, i.e materials having an isotropic structure,
often have anisotropic fracture surfaces only because of the
breaking of isotropy by the initial conditions. In such cases it
appears useful to analyze the anisotropic contributions as
``corrections to scaling" beyond the isotropic sector, which is
always there, with a leading scaling exponent. The analysis reveals
non-universality in the scaling exponents, a finding that calls for
further future study and assessment, including the interesting
question of the possible existence of universality classes. On the
practical side, we have demonstrated that the full information
concerning the two dimensional structure function can be efficiently
parameterized by a few amplitudes and scaling exponents. The reader
should note that this Letter dealt only with second order structure
functions. In analogy to turbulence it may be possible to decompose
any higher order structure function into SO(2) irreducible
representations \cite{99ALP}. This may reveal additional interesting
scaling properties such as the phenomenon of multiscaling
\cite{05BouP}. Finally, we would like to emphasize the great
interest in the proposed isotropic fracture experiment and the
measurement of the roughness exponent in such an experiment. If
indeed this scaling exponent were identical to the exponent of the
isotropic sector in a standard experiment, this would significantly
strengthen the theoretical interest in the approach proposed in this
Letter.

We are grateful to D.Bonamy, E.Bouchaud, J.Fineberg and A.Sagy for
sharing with us their valuable experimental fracture surface data.
We acknowledge useful discussions with D. Bonamy. E.B. is supported
by the Horowitz Complexity Science Foundation. This work was
supported in part by the Israel Science Foundation administered by
the Israel Academy of Sciences.

\end{document}